\documentclass[eqsecnum,preprint,prd,aps,nofootinbib]{revtex4}
\usepackage{amsmath}
\usepackage{psfig}
\begin{document}
\title{Spectral asymptotics of Euclidean quantum gravity 
with diff-invariant boundary conditions}
\author{Giampiero Esposito$^{1,2}$
\thanks{Electronic address: giampiero.esposito@na.infn.it}
Guglielmo Fucci$^{2}$
\thanks{Electronic address: guglielmo.fucci@na.infn.it}
Alexander Yu. Kamenshchik$^{3,4}$
\thanks{Electronic address: sasha.kamenshchik@centrovolta.it}
Klaus Kirsten$^{5}$
\thanks{Electronic address: Klaus_Kirsten@baylor.edu}}
\address{${ }^{1}$Istituto Nazionale di 
Fisica Nucleare, Sezione di Napoli,\\
Complesso Universitario di Monte S. Angelo, Via Cintia,
Edificio N', 80126 Napoli, Italy\\
${ }^{2}$Dipartimento di Scienze Fisiche,
Complesso Universitario di Monte S. Angelo,\\
Via Cintia, Edificio N', 80126 Napoli, Italy\\
${ }^{3}$Dipartimento di Scienze Fisiche e Matematiche, 
Universit\`a dell'Insubria, Como, Italy\\
${ }^{4}$L.D. Landau Institute for Theoretical Physics of the
Russian Academy of Sciences, Moscow, Russia\\
${ }^{5}$Department of Mathematics, Baylor University, Waco
TX 76798, USA}

\vspace{2cm}
\begin{abstract}
A general method is known to exist for studying Abelian and 
non-Abelian gauge theories, as well as Euclidean quantum gravity,
at one-loop level on manifolds with boundary. In the latter case,
boundary conditions on metric perturbations $h$ can be chosen to
be completely invariant under infinitesimal diffeomorphisms, to
preserve the invariance group of the theory and BRST symmetry. In 
the de Donder gauge, however, the resulting boundary-value problem for
the Laplace type operator acting on $h$ is 
known to be self-adjoint but not strongly
elliptic. The latter is a technical condition ensuring that 
a unique smooth solution of the boundary-value problem exists,
which implies, in turn, that the global heat-kernel 
asymptotics yielding one-loop divergences and one-loop effective
action actually exists. 
The present paper shows that, on the Euclidean four-ball,
only the scalar part of perturbative modes
for quantum gravity are affected by the lack of strong ellipticity.
Further evidence for lack of strong ellipticity, from an analytic
point of view, is therefore obtained. Interestingly, three sectors of the
scalar-perturbation problem remain elliptic, while lack of strong
ellipticity is ``confined'' to the remaining fourth sector.
The integral representation of the resulting $\zeta$-function 
asymptotics is also obtained; this 
remains regular at the origin by virtue 
of a spectral identity here obtained for the first time.

\end{abstract}
\maketitle
\section{Introduction}
The study of gauge theories and quantum gravity on manifolds with
boundary is motivated by the problems of quantum 
cosmology \cite{[1]} and quantum field
theory under the influence of external conditions \cite{[2]},
and several efforts have been produced in the literature to study
boundary conditions and one-loop semiclassical properties within 
this framework \cite{[3]}. In our paper we are interested in boundary 
conditions for metric perturbations that are completely invariant 
under infinitesimal diffeomorphisms, since they are part of the
general scheme according to which the boundary conditions are preserved
under the action of the symmetry group of the theory 
\cite{[4],[5],[6]}.
In field-theoretical language, this means setting to zero at the
boundary that part $\pi A$ of the gauge field $A$ that lives on
the boundary ${\cal B}$ ($\pi$ being a projection operator):
\begin{equation}
\Bigr[\pi A \Bigr]_{\cal B}=0,
\label{(1.1)}
\end{equation}
as well as the gauge-fixing functional,
\begin{equation}
\Bigr[\Phi(A)\Bigr]_{\cal B}=0,
\label{(1.2)}
\end{equation}
and the whole ghost field
\begin{equation}
[\varphi]_{\cal B}=0.
\label{(1.3)}
\end{equation}

For Euclidean quantum gravity, Eq. (1.1) reads as
\begin{equation}
[h_{ij}]_{\cal B}=0,
\label{(1.4)}
\end{equation}
where $h_{ij}$ are perturbations of the induced three-metric. 
To arrive at the gravitational 
counterpart of Eqs. (1.2) and (1.3), note first that, 
under infinitesimal diffeomorphisms, metric
perturbations $h_{\mu \nu}$ transform according to
\begin{equation}
{\widehat h}_{\mu \nu} \equiv h_{\mu \nu}+
\nabla_{(\mu} \; \varphi_{\nu)},
\label{(1.5)}
\end{equation}
where $\nabla$ is the Levi--Civita connection on the background
four-geometry with metric $g$, and $\varphi_{\nu}dx^{\nu}$ is
the ghost one-form (strictly, our presentation is simplified:
there are two independent ghost fields obeying Fermi statistics,
and we will eventually multiply by $-2$ the effect of $\varphi_{\nu}$
to take this into account). In geometric language, the infinitesimal
variation $\delta h_{\mu \nu} \equiv {\widehat h}_{\mu \nu}
-h_{\mu \nu}$ is given by the Lie derivative along $\varphi$ of the
four-metric $g$. For manifolds with boundary, 
Eq. (1.5) implies that \cite{[7],[8]}
\begin{equation}
{\widehat h}_{ij}=h_{ij}+\varphi_{(i \mid j)}
+K_{ij}\varphi_{0},
\label{(1.6)}
\end{equation}
where the stroke denotes three-dimensional covariant differentiation
tangentially with respect to the intrinsic Levi--Civita connection 
of the boundary, while $K_{ij}$ is the extrinsic-curvature tensor of
the boundary. Of course, $\varphi_{0}$ and $\varphi_{i}$ are the 
normal and tangential components of the ghost, respectively. By virtue
of Eq. (1.6), the boundary conditions (1.4) are ``gauge invariant'', i.e.
\begin{equation}
\Bigr[{\widehat h}_{ij}\Bigr]_{\cal B}=0,
\label{(1.7)}
\end{equation}
if and only if the whole ghost field obeys homogeneous Dirichlet
conditions, so that
\begin{equation}
[\varphi_{0}]_{\cal B}=0,
\label{(1.8)}
\end{equation}
\begin{equation}
[\varphi_{i}]_{\cal B}=0.
\label{(1.9)}
\end{equation}
The conditions (1.8) and (1.9) are necessary and sufficient since 
$\varphi_{0}$ and $\varphi_{i}$ are independent, and three-dimensional
covariant differentiation commutes with the operation of restriction
to the boundary. We are indeed assuming that the boundary 
$\cal B$ is smooth and not totally geodesic, i.e.
$K_{ij} \not = 0$. However, for totally geodesic manifolds, having
$K_{ij}=0$, the condition (1.8) is no longer necessary. 

On imposing boundary conditions on the remaining set of metric
perturbations, the key point is to make sure that {\it the invariance
of such boundary conditions under the infinitesimal transformations}
(1.5) {\it is again guaranteed by} (1.8) {\it and} (1.9), since otherwise
one would obtain incompatible sets of boundary conditions on the
ghost field. Indeed, on using the DeWitt--Faddeev--Popov formalism
for the $\langle {\rm out}| {\rm in} \rangle$ amplitudes of quantum
gravity, it is necessary to use a gauge-averaging term in the
Euclidean action, of the form \cite{[9]}
\begin{equation}
I_{g.a.}={1\over 16 \pi G}\int_{{\cal M}}
{\Phi_{\nu}\Phi^{\nu}\over 2\alpha}\sqrt{{\rm det} \; g} \; d^{4}x,
\label{(1.10)}
\end{equation}
where $\Phi_{\nu}$ is any functional which leads to self-adjoint 
(elliptic) operators on metric and ghost perturbations. One then
finds that
\begin{equation}
\delta \Phi_{\mu}(h) \equiv \Phi_{\mu}(h)-
\Phi_{\mu}({\widehat h})={\cal F}_{\mu}^{\; \nu} \;
\varphi_{\nu},
\label{(1.11)}
\end{equation}
where ${\cal F}_{\mu}^{\; \nu}$ is an elliptic operator that acts
linearly on the ghost field. Thus, if one imposes the boundary
conditions
\begin{equation}
\Bigr[\Phi_{\mu}(h)\Bigr]_{\cal B}=0,
\label{(1.12)}
\end{equation}
and if one assumes that the ghost field can be expanded in a
complete orthonormal set of eigenfunctions $u_{\nu}^{\; (\lambda)}$
of ${\cal F}_{\mu}^{\; \nu}$ which vanish at the boundary, i.e.
\begin{equation}
{\cal F}_{\mu}^{\; \nu} u_{\nu}^{\; (\lambda)}=\lambda
u_{\mu}^{\; (\lambda)},
\label{(1.13)}
\end{equation}
\begin{equation}
\varphi_{\nu}=\sum_{\lambda}C_{\lambda}u_{\nu}^{\; (\lambda)},
\label{(1.14)}
\end{equation}
\begin{equation}
\Bigr[u_{\mu}^{\; (\lambda)}\Bigr]_{\cal B}=0,
\label{(1.15)}
\end{equation}
the boundary conditions (1.12) are automatically gauge-invariant under
the Dirichlet conditions (1.8) and (1.9) on the ghost.

Having obtained the general recipe expressed by Eqs. (1.4) and (1.12), we
can recall what they imply on the Euclidean four-ball. This background
is relevant for one-loop quantum cosmology in the limit of small
three-geometry on the one hand \cite{[10]}, 
and for spectral geometry and spectral
asymptotics on the other hand \cite{[11]}. As shown in Ref. \cite{[7]}, 
if one chooses the de Donder gauge-fixing functional
\begin{equation}
\Phi_{\mu}(h)=\nabla^{\nu}\Bigr(h_{\mu \nu}-{1\over 2}
g_{\mu \nu}g^{\rho \sigma}h_{\rho \sigma}\Bigr),
\label{(1.16)}
\end{equation}
which has the virtue of leading to an operator of Laplace type on
$h_{\mu \nu}$ in the one-loop functional integral, Eq. (1.12) yields 
the mixed boundary conditions
\begin{equation}
\left[{\partial h_{00}\over \partial \tau}+{6\over \tau}h_{00}
-{\partial \over \partial \tau}(g^{ij}h_{ij})
+{2\over \tau^{2}}h_{0i}^{\; \mid i}\right]_{\cal B}=0,
\label{(1.17)}
\end{equation}
\begin{equation}
\left[{\partial h_{0i}\over \partial \tau}+{3\over \tau}h_{0i}
-{1\over 2}{\partial h_{00}\over \partial x^{i}}
\right]_{\cal B}=0.
\label{(1.18)}
\end{equation}
In Refs. \cite{[3],[7]}, 
the boundary conditions (1.4), (1.17) and (1.18) were 
used to evaluate the full one-loop divergence of quantized general
relativity on the Euclidean four-ball, including all $h_{\mu \nu}$
and all ghost modes. However, the meaning of such a calculation
became unclear after the discovery in Ref. \cite{[6]} that the 
boundary-value problem for the Laplacian $P$ acting on metric
perturbations is not strongly elliptic by virtue of tangential
derivatives in the boundary conditions (1.17) and (1.18). Strong
ellipticity \cite{[11]} is a technical requirement ensuring that a unique
smooth solution of the boundary-value problem exists which vanishes
at infinite geodesic distance from the boundary (see Appendix A). 
If it is fulfilled,
this ensures that the $L^{2}$ trace of the heat semigroup $e^{-tP}$ 
exists, with the associated global heat-kernel asymptotics that 
yields one-loop divergence and one-loop effective action. However,
when strong ellipticity does not hold, the $L^{2}$ trace of
$e^{-tP}$ acquires a singular part \cite{[6]}, and hence it is unclear how
to attach a meaning to $\zeta$-function calculations. 

All of this has motivated our analysis, which therefore starts in Sec.
II with the mode-by-mode form of the boundary conditions (1.4), (1.8), 
(1.9), (1.17) and (1.18) with the resulting eigenvalue conditions.
Section III studies the matrix for coupled scalar modes, while Sec. IV
obtains the first pair of resulting scalar-mode 
$\zeta$-functions and Sec. V studies the remaining
elliptic and non-elliptic parts of spectral asymptotics. Results and
open problems are described in Sec. VI, while technical details are
given in the Appendices. 

\section{Eigenvalue conditions on the four-ball}

On the Euclidean four-ball, which can be viewed as the portion of
flat Euclidean four-space bounded by a three-sphere of radius $q$,
metric perturbations $h_{\mu \nu}$ can be
expanded in terms of hyperspherical harmonics as \cite{[12], [13]}
\begin{equation}
h_{00}(x,\tau)=\sum_{n=1}^{\infty}a_{n}(\tau)Q^{(n)}(x),
\label{(2.1)}
\end{equation}
\begin{equation}
h_{0i}(x,\tau)=\sum_{n=2}^{\infty}\left[b_{n}(\tau)
{Q_{\mid i}^{(n)}(x)\over (n^{2}-1)}+c_{n}(\tau)S_{i}^{(n)}(x)\right],
\label{(2.2)}
\end{equation}
\begin{eqnarray}
h_{ij}(x,\tau)&=& \sum_{n=3}^{\infty}d_{n}(\tau)\left[
{Q_{\mid ij}^{(n)}(x)\over (n^{2}-1)}+{c_{ij}\over 3}
Q^{(n)}(x)\right]+\sum_{n=1}^{\infty}{e_{n}(\tau)\over 3}c_{ij}
Q^{(n)}(x) \nonumber \\
&+& \sum_{n=3}^{\infty}\left[f_{n}(\tau)\Bigr(S_{i \mid j}^{(n)}(x)
+S_{j \mid i}^{(n)}(x) \Bigr)+k_{n}(\tau)G_{ij}^{(n)}(x)\right],
\label{(2.3)}
\end{eqnarray}
where $\tau \in [0,q]$ and 
$Q^{(n)}(x), S_{i}^{(n)}(x)$ and $G_{ij}^{(n)}(x)$ are scalar,
transverse vector and transverse-traceless tensor hyperspherical
harmonics, respectively, on a unit three-sphere with metric $c_{ij}$. 
By insertion of the expansions (2.1)-(2.3) into the eigenvalue equation
for the Laplacian acting on $h_{\mu \nu}$, and by setting 
$\sqrt{E} \rightarrow iM$, which corresponds to a rotation of
contour in the $\zeta$-function analysis \cite{[14]}, 
one finds the modes as
linear combinations of modified Bessel functions of first kind 
according to \cite{[13]}
\begin{equation}
a_{n}(\tau)={1\over \tau}\Bigr[\gamma_{1}I_{n}(M\tau)
+\gamma_{3}I_{n-2}(M\tau)+\gamma_{4}I_{n+2}(M\tau)\Bigr],
\label{(2.4)}
\end{equation}
\begin{equation}
b_{n}(\tau)=\gamma_{2}I_{n}(M\tau)+(n+1)\gamma_{3}I_{n-2}(M\tau)
-(n-1)\gamma_{4}I_{n+2}(M\tau),
\label{(2.5)}
\end{equation}
\begin{equation}
c_{n}(\tau)=\varepsilon_{1}I_{n+1}(M\tau)+\varepsilon_{2}I_{n-1}(M\tau),
\label{(2.6)}
\end{equation}
\begin{equation}
d_{n}(\tau)=\tau \Bigr[-\gamma_{2}I_{n}(M\tau)+{(n+1)\over (n-2)}
\gamma_{3}I_{n-2}(M\tau)+{(n-1)\over (n+2)}\gamma_{4}I_{n+2}(M\tau)
\Bigr],
\label{(2.7)}
\end{equation}
\begin{equation}
e_{n}(\tau)=\tau \Bigr[(3\gamma_{1}-2 \gamma_{2})I_{n}(M\tau)
-\gamma_{3}I_{n-2}(M\tau)-\gamma_{4}I_{n+2}(M\tau)\Bigr],
\label{(2.8)}
\end{equation}
\begin{equation}
f_{n}(\tau)=\tau \left[-{\varepsilon_{1}\over (n+2)}I_{n+1}(M\tau)
+{\varepsilon_{2}\over (n-2)}I_{n-1}(M\tau)\right],
\label{(2.9)}
\end{equation}
\begin{equation}
k_{n}(\tau)=\alpha_{1}\tau I_{n}(M\tau).
\label{(2.10)}
\end{equation}
Modified Bessel functions of second kind are not included to ensure
regularity at the origin $\tau=0$.
Moreover, normal and tangential components of the ghost field admit the
following expansion on the four-ball:
\begin{equation}
\varphi_{0}(x,\tau)=\sum_{n=1}^{\infty}l_{n}(\tau)Q^{(n)}(x),
\label{(2.11)}
\end{equation}
\begin{equation}
\varphi_{i}(x,\tau)=\sum_{n=2}^{\infty}\left[m_{n}(\tau)
{Q_{\mid i}^{(n)}(x)\over (n^{2}-1)}+p_{n}(\tau)S_{i}^{(n)}(x)
\right],
\label{(2.12)}
\end{equation}
where the ghost modes $l_{n}(\tau),m_{n}(\tau)$ and $p_{n}(\tau)$ are
found to read as \cite{[13]}
\begin{equation}
l_{n}(\tau)={1\over \tau}\Bigr[\kappa_{1}I_{n+1}(M\tau)
+\kappa_{2}I_{n-1}(M\tau)\Bigr],
\label{(2.13)}
\end{equation}
\begin{equation}
m_{n}(\tau)=-(n-1)\kappa_{1}I_{n+1}(M\tau)+(n+1)\kappa_{2}
I_{n-1}(M\tau),
\label{(2.14)}
\end{equation}
\begin{equation}
p_{n}(\tau)=\theta I_{n}(M\tau).
\label{(2.15)}
\end{equation}
At this stage, the boundary conditions (1.4), (1.17), (1.18), (1.8) and 
(1.9) can be re-expressed in terms of metric and ghost modes as
\begin{equation}
{da_{n}\over d\tau}+{6\over \tau}a_{n}-{1\over \tau^{2}}
{de_{n}\over d\tau}-{2\over \tau^{2}}b_{n}=0 \; \; {\rm on}
\; S^{3},
\label{(2.16)}
\end{equation}
\begin{equation}
{db_{n}\over d\tau}+{3\over \tau}b_{n}-{(n^{2}-1)\over 2}a_{n}=0
\; \; {\rm on} \; S^{3},
\label{(2.17)}
\end{equation}
\begin{equation}
{dc_{n}\over d\tau}+{3\over \tau}c_{n}=0
\; \; {\rm on} \; S^{3},
\label{(2.18)}
\end{equation}
\begin{equation}
d_{n}=e_{n}=f_{n}=k_{n}=l_{n}=m_{n}=p_{n}=0 \; \; {\rm on} \;
S^{3}.
\label{(2.19)}
\end{equation}
Furthermore, the formulae (2.4)--(2.10) and (2.13)--(2.15) can be used to
obtain homogeneous linear systems that yield, implicitly, the
eigenvalues of our problem. The conditions for finding non-trivial
solutions of such linear systems are given by the vanishing of the
associated determinants; these yield the eigenvalue conditions
$\delta(E)=0$, i.e. the equations obeyed by the eigenvalues by
virtue of the boundary conditions. For the purpose of a
rigorous analysis, we need the full expression of such
eigenvalue conditions for each set of coupled modes. 
Upon setting $\sqrt{E} \rightarrow iM$, we denote by $D(Mq)$ the
counterpart of $\delta(E)$, bearing in mind that, strictly,
only $\delta(E)$ yields implicitly the eigenvalues, while
$D(Mq)$ is more convenient for $\zeta$-function calculations
\cite{[14]}.

To begin, the decoupled vector mode $c_{2}(\tau)=I_{3}(M\tau)$ obeys the
Robin boundary condition (2.18), which yields 
\begin{equation}
D(Mq)=I_{2}(Mq)+I_{4}(Mq)+{6\over Mq}I_{3}(Mq),
\label{(2.20)}
\end{equation}
with degeneracy $6$. Coupled vector modes $c_{n}(\tau)$ and
$f_{n}(\tau)$ obey the boundary conditions (2.18) and (2.19), and hence
the corresponding $D(Mq)$ reads as
\begin{eqnarray}
D_{n}(Mq)& = & I_{n-1}(Mq)\Bigr(I_{n}(Mq)+I_{n+2}(Mq)+{6\over Mq}
I_{n+1}(Mq)\Bigr) \nonumber \\
&+& {(n-2)\over (n+2)}I_{n+1}(Mq)\Bigr(I_{n-2}(Mq)+I_{n}(Mq)
+{6\over Mq}I_{n-1}(Mq)\Bigr),
\label{(2.21)}
\end{eqnarray}
with degeneracy $2(n^{2}-1)$, for all $n \geq 3$.

The scalar modes
\begin{equation}
a_{1}(\tau)={1\over \tau}\Bigr(\gamma_{1}I_{1}(M\tau)
+\gamma_{4}I_{3}(M\tau)\Bigr),
\label{(2.22)}
\end{equation}
\begin{equation}
e_{1}(\tau)=\tau \Bigr(3\gamma_{1}I_{1}(M\tau)
-\gamma_{4}I_{3}(M\tau)\Bigr),
\label{(2.23)}
\end{equation}
obey the boundary conditions
\begin{equation}
{da_{1}\over d\tau}+{6\over \tau}-{1\over \tau^{2}}
{de_{1}\over d\tau}=0 \; \; {\rm at} \; \; \tau=q,
\label{(2.24)}
\end{equation}
\begin{equation}
e_{1}(q)=0,
\label{(2.25)}
\end{equation}
which imply
\begin{eqnarray}
D(Mq) & = & 20 I_{1}(Mq)I_{3}(Mq)-Mq(I_{0}(Mq)+I_{2}(Mq))I_{3}(Mq)
\nonumber \\
&+& 3Mq I_{1}(Mq)(I_{2}(Mq)+I_{4}(Mq)),
\label{(2.26)}
\end{eqnarray}
with degeneracy $1$.

The scalar modes
\begin{equation}
a_{2}(\tau)={1\over \tau}\Bigr[\gamma_{1}I_{2}(M\tau)
+\gamma_{4}I_{4}(M\tau)\Bigr],
\label{(2.27)}
\end{equation}
\begin{equation}
b_{2}(\tau)=\gamma_{2}I_{2}(M\tau)-\gamma_{4}I_{4}(M\tau),
\label{(2.28)}
\end{equation}
\begin{equation}
e_{2}(\tau)=\tau \Bigr[(3\gamma_{1}-2\gamma_{2})I_{2}(M\tau)
-\gamma_{4}I_{4}(M\tau)\Bigr],
\label{(2.29)}
\end{equation}
obey the boundary conditions (2.16), (2.17) and (2.19) with $n=2$, and
hence yield the determinant
\begin{equation}
D(Mq)={\rm det} 
\begin{pmatrix}
I_{2}(Mq)-MqI_{2}'(Mq) \hfill & MqI_{2}'(Mq) \hfill & 
4I_{4}(Mq)+Mq I_{4}'(Mq) \hfill \\
3I_{2}(Mq) \hfill & -(2Mq I_{2}'(Mq)+6I_{2}(Mq)) \hfill &
2MqI_{4}'(Mq)+9I_{4}(Mq) \hfill \\
3I_{2}(Mq) \hfill & -2I_{2}(Mq) \hfill & -I_{4}(Mq) \hfill
\end{pmatrix},
\label{(2.30)}
\end{equation}
with degeneracy $4$.

For all $n \geq 3$, coupled scalar modes $a_{n},b_{n},d_{n},e_{n}$ 
obey the boundary conditions (2.16), (2.17), (2.19). The resulting 
determinant reads as
\begin{equation}
D_{n}(Mq)={\rm det} \; \rho_{ij}(Mq),
\label{(2.31)}
\end{equation}
with degeneracy $n^{2}$, where $\rho_{ij}$ is a $4 \times 4$ 
matrix with entries
\begin{equation}
\rho_{11}=I_{n}(Mq)-Mq I_{n}'(Mq), \;
\rho_{12}=Mq I_{n}'(Mq),
\label{(2.32)}
\end{equation}
\begin{equation}
\rho_{13}=(2-n)I_{n-2}(Mq)+MqI_{n-2}'(Mq), \;
\rho_{14}=(2+n)I_{n+2}(Mq)+MqI_{n+2}'(Mq),
\label{(2.33)}
\end{equation}
\begin{equation}
\rho_{21}=-(n^{2}-1)I_{n}(Mq), \;
\rho_{22}=2MqI_{n}'(Mq)+6I_{n}(Mq),
\label{(2.34)}
\end{equation}
\begin{equation}
\rho_{23}=2(n+1)MqI_{n-2}'(Mq)-(n^{2}-6n-7)I_{n-2}(Mq),
\label{(2.35)}
\end{equation}
\begin{equation}
\rho_{24}=-2(n-1)MqI_{n+2}'(Mq)-(n^{2}+6n-7)I_{n+2}(Mq),
\label{(2.36)}
\end{equation}
\begin{equation}
\rho_{31}=0, \;
\rho_{32}=-I_{n}(Mq),
\label{(2.37)}
\end{equation}
\begin{equation}
\rho_{33}={(n+1)\over (n-2)}I_{n-2}(Mq), \;
\rho_{34}={(n-1)\over (n+2)}I_{n+2}(Mq),
\label{(2.38)}
\end{equation}
\begin{equation}
\rho_{41}=3I_{n}(Mq), \;
\rho_{42}=-2I_{n}(Mq), \;
\rho_{43}=-I_{n-2}(Mq), \;
\rho_{44}=-I_{n+2}(Mq).
\label{(2.39)}
\end{equation}

Transverse-traceless tensor modes $k_{n}(\tau)$ yield, by virtue
of Eqs. (2.10) and (2.19), 
\begin{equation}
D_{n}(Mq)=I_{n}(Mq), \; \forall n \geq 3,
\label{(2.40)}
\end{equation}
with degeneracy $2(n^{2}-4)$. 

As far as ghost modes are concerned, the decoupled mode 
$l_{1}(\tau)={1\over \tau}I_{2}(M\tau)$ vanishes at the three-sphere
boundary and hence yields 
\begin{equation}
D(Mq)=I_{2}(Mq),
\label{(2.41)}
\end{equation}
with degeneracy $1$, while scalar and vector ghost modes lead to
\begin{equation}
D_{n}(Mq)=I_{n+1}(Mq), \; \; \forall n \geq 2,
\label{(2.42)}
\end{equation}
and
\begin{equation}
D_{n}(Mq)=I_{n}(Mq), \; \; \forall n \geq 2,
\label{(2.43)}
\end{equation}
respectively, with degeneracy $n^{2}$ for Eq. (2.42) and
$2(n^{2}-1)$ for Eq. (2.43).

Our $D(Mq)$ equations can be re-expressed in a very helpful way by
using repeatedly the identities for modified Bessel
functions and their derivatives in the Appendix. Hence we find,
on setting $w \equiv Mq$, that
\begin{equation}
{D(w)\over 2}=I_{2}(w)
\label{(2.44)}
\end{equation}
for the decoupled vector mode in Eq. (2.20), while coupled vector
modes in Eq. (2.21) yield
\begin{equation}
{(n+2)\over 4n}D_{n}(w)=I_{n}(w)\left(I_{n}'(w)+{2 \over w}I_{n}(w)\right).
\label{(2.45)}
\end{equation}
Moreover, the scalar modes $a_{1},e_{1}$ ruled by Eq. (2.26) yield
\begin{equation}
{D(w)\over 4w}=I_{2}(w)\left(I_{2}'(w)+{4\over w}I_{2}(w)\right),
\label{(2.46)}
\end{equation}
and the scalar modes $a_{2},b_{2},e_{2}$ ruled by Eq. (2.30) lead to
\begin{equation}
{D(w)\over 6w^{2}}=I_{1}(w)I_{3}(w)
\left(I_{3}'(w)+{5 \over w}I_{3}(w)\right).
\label{(2.47)}
\end{equation}

\section{Matrix for coupled scalar modes}

The hardest part of our analysis is the investigation of Eq. (2.31).
For this purpose, we first exploit the formulae in Appendix B
to find
\begin{equation}
\rho_{11}=I_{n}(w)-wI_{n}'(w), \; 
\rho_{12}=wI_{n}'(w), \; 
\rho_{13}=wI_{n}'(w)+nI_{n}(w), \;
\rho_{14}=wI_{n}'(w)-nI_{n}(w),
\label{(3.1)}
\end{equation}
\begin{equation}
\rho_{21}=-(n^{2}-1)I_{n}(w), \;
\rho_{22}=2(wI_{n}'(w)+3I_{n}(w)),
\label{(3.2)}
\end{equation}
\begin{equation}
\rho_{23}=(n+1)\biggr \{ \left[3(n+1)+{2n(n-1)(n+3)\over w^{2}}\right]
I_{n}(w) 
+ 2 \left[w+{(n-1)(n+3)\over w}\right]I_{n}'(w) \biggr \},
\label{(3.3)}
\end{equation}
\begin{equation}
\rho_{24}=
(n-1)\biggr \{ \left[3(n-1)+{2n(n+1)(n-3)\over w^{2}}\right]
I_{n}(w) 
- 2 \left[w+{(n+1)(n-3)\over w}\right]I_{n}'(w) \biggr \},
\label{(3.4)}
\end{equation}
\begin{equation}
\rho_{31}=0, \; \rho_{32}=-I_{n}(w),
\label{(3.5)}
\end{equation}
\begin{equation}
\rho_{33}={(n+1)\over (n-2)}\left[\left(1+{2n(n-1)\over w^{2}}
\right)I_{n}(w)+{2(n-1)\over w}I_{n}'(w)\right],
\label{(3.6)}
\end{equation}
\begin{equation}
\rho_{34}={(n-1)\over (n+2)}\left[\left(1+{2n(n+1)\over w^{2}}
\right)I_{n}(w)-{2(n+1)\over w}I_{n}'(w)\right],
\label{(3.7)}
\end{equation}
\begin{equation}
\rho_{41}=3I_{n}(w), \; \rho_{42}=-2I_{n}(w),
\label{(3.8)}
\end{equation}
\begin{equation}
\rho_{43}=-\left(1+{2n(n-1)\over w^{2}}\right)I_{n}(w)
-{2(n-1)\over w}I_{n}'(w),
\label{(3.9)}
\end{equation}
\begin{equation}
\rho_{44}=-\left(1+{2n(n+1)\over w^{2}}\right)I_{n}(w)
+{2(n+1)\over w}I_{n}'(w).
\label{(3.10)}
\end{equation}
The resulting determinant, despite its cumbersome expression, 
can be studied by introducing the variable
\begin{equation}
y \equiv {I_{n}'(w)\over I_{n}(w)},
\label{(3.11)}
\end{equation}
which leads to 
\begin{equation}
D_{n}(w)={48n(1-n^{2})\over (n^{2}-4)}
I_{n}^{4}(w)(y-y_{1})(y-y_{2})(y-y_{3})(y-y_{4}),
\label{(3.12)}
\end{equation}
where
\begin{equation}
y_{1} \equiv -{n\over w}, \;
y_{2} \equiv {n\over w}, \;
y_{3} \equiv -{n \over w}-{w \over 2}, \;
y_{4} \equiv {n \over w}-{w \over 2},
\label{(3.13)}
\end{equation}
and hence
\begin{eqnarray}
{(n^{2}-4)\over 48n(1-n^{2})}D_{n}(w)
&=&\left(I_{n}'(w)+{n \over w}I_{n}(w)\right)
\left(I_{n}'(w)-{n \over w}I_{n}(w)\right) \nonumber \\
& \times & \left(I_{n}'(w)+\left({w \over 2}+{n \over w}
\right)I_{n}(w)\right)
\left(I_{n}'(w)+\left({w \over 2}-{n \over w}
\right)I_{n}(w)\right).
\label{(3.14)}
\end{eqnarray}

\section{First pair of scalar-mode $\zeta$-functions}

Equations (2.40)--(2.47) and (3.14) are sufficient to obtain an integral
representation of the $\zeta$-function, the residues of which yield
all heat-kernel coefficients. This topic is described in great detail
in the existing literature (see, for example, Refs. [11] and [15]) and
hence we limit ourselves to a very brief outline before presenting our
results. 

Given the elliptic operator $P$ acting on physical fields defined on
the $m$-dimensional Riemannian manifold $\cal M$, one can build the
associated heat kernel $U(x,y;t)$ and the corresponding integrated
heat kernel (bundle indices are not written down explicitly, but the
fibre trace ${\rm tr}$ takes them into account)
\begin{equation}
{\rm Tr}_{L^{2}}e^{-tP}=\int_{{\cal M}}{\rm tr}U(x,x;t)\sqrt{g}\; d^{m}x,
\label{(4.1)}
\end{equation}
which has the asymptotic expansion, as $t \rightarrow 0^{+}$,
\begin{equation}
{\rm Tr}_{L^{2}}e^{-tP} \sim (4 \pi t)^{-{m\over 2}}
\sum_{k=0}^{\infty}A_{{k\over 2}}t^{{k\over 2}}.
\label{(4.2)}
\end{equation}
The $A_{{k\over 2}}$ coefficients are said to describe the global
asymptotics in that they are obtained by 
integration over $\cal M$ and its
boundary $\cal B$ of local geometric invariants built from the
Riemann curvature of $\cal M$, gauge curvature, extrinsic curvature of
$\cal B$, potential terms in $P$ and in the boundary operator 
expressing the boundary conditions. On the other hand,
since the $\zeta$-function of 
$P$ is related to the integrated heat kernel by an inverse Mellin
transform [1,3,11]:
\begin{equation}
\zeta_{P}(s)={\rm Tr}_{L^{2}}P^{-s}=\sum_{\lambda_{P}}\lambda_{P}^{-s}
={1\over \Gamma(s)}\int_{0}^{\infty}t^{s-1}\Bigr({\rm Tr}_{L^{2}}
e^{-tP}\Bigr)dt,
\label{(4.3)}
\end{equation}
the global heat-kernel coefficients in the asymptotic expansion (4.2)
can be also obtained from the residues of $\zeta_{P}(s)$ [15].

Moreover, since the function occurring in the equation obeyed by
the eigenvalues by virtue of the boundary conditions admits a
canonical-product representation [1,3], one can also express 
$\zeta_{P}(s)$ as a contour integral which is eventually rotated
to the imaginary axis. The residues of the latter integral yield
therefore the $A_{{k\over 2}}$ coefficients used in evaluating 
one-loop effective action and one-loop divergences.

In our problem the $P$ operator is the Laplacian on the Euclidean 
four-ball acting on metric perturbations. Equations (2.40)--(2.47) 
correspond to a familiar mixture of Dirichlet and Robin boundary
conditions for which integral representation of the
$\zeta$-function and heat-kernel coefficients are immediately obtained.
New features arise instead from Eq. (3.14), that gives rise,
at first sight, to four different $\zeta$-functions.
On studying the first line of Eq. (3.14), we
can exploit the work in Ref. [16] and the uniform asymptotic 
expansion of Bessel functions and their first derivatives (see
Appendix B) to say that the integral representation of the resulting
$\zeta$-function reads as
\begin{equation}
\zeta_{A}^{\pm}(s) \equiv {(\sin \pi s)\over \pi}\sum_{n=3}^{\infty}
n^{-(2s-2)}  
\int_{0}^{\infty}dz \; z^{-2s}{\partial \over \partial z}
{\rm log}\left[z^{-\beta_{\pm}(n)}\Bigr(znI_{n}'(zn) 
\pm n I_{n}(zn)\Bigr)\right].
\label{(4.4)}
\end{equation}
With our notation, $\beta_{+}(n)=n, \beta_{-}(n)=n+2$, where these
factors are fixed by the leading behaviour of the eigenvalue
condition as $z \rightarrow 0$ \cite{[15]};
the uniform asymptotic expansion of modified Bessel functions
and their first derivatives (see Appendix B) can be used to find
(hereafter $\tau=\tau(z) \equiv (1+z^{2})^{-{1\over 2}}$
from Eq. (B8))
\begin{equation}
znI_{n}'(zn) \pm n I_{n}(zn) \sim
{n \over \sqrt{2\pi n}}{e^{n \eta}\over \sqrt{\tau}}
(1 \pm \tau)\left(1+\sum_{k=1}^{\infty}{p_{k,\pm}(\tau)\over n^{k}}
\right),
\label{(4.5)}
\end{equation}
where
\begin{equation}
p_{k,\pm}(\tau) \equiv (1\pm \tau)^{-1}\Bigr(v_{k}(\tau) \pm 
\tau u_{k}(\tau)\Bigr),
\label{(4.6)}
\end{equation}
for all $k \geq 1$, and
\begin{equation}
{\rm log} \left(1+\sum_{k=1}^{\infty}
{p_{k,\pm}(\tau)\over n^{k}}\right) \sim
\sum_{k=1}^{\infty}{T_{k,\pm}(\tau)\over n^{k}}.
\label{(4.7)}
\end{equation}
Thus, the $\zeta$-functions (4.4) obtain, from the first pair of 
round brackets in Eq. (4.5), the contributions (cf. Ref. \cite{[16]})
\begin{equation}
A_{+}(s) \equiv 
\sum_{n=3}^{\infty} n^{-(2s-2)} 
{(\sin \pi s)\over \pi}
\int_{0}^{\infty}dz \; z^{-2s}{\partial \over \partial z}
\log \Bigr(1 + (1+z^{2})^{-{1\over 2}}\Bigr),
\label{(4.8)}
\end{equation}
\begin{equation}
A_{-}(s) \equiv 
\sum_{n=3}^{\infty} n^{-(2s-2)} 
{(\sin \pi s)\over \pi}
\int_{0}^{\infty}dz \; z^{-2s}{\partial \over \partial z}
\log \left({1 - (1+z^{2})^{-{1\over 2}} \over z^{2}}\right),
\label{(4.9)}
\end{equation}
where $z^{2}$ in the denominator of the argument of the $\log$
arises, in Eq. (4.9), from the extra $z^{-2}$ in the prefactor
$z^{-\beta_{-}(n)}$ in the definition (4.4). Moreover, the
second pair of round brackets in Eq. (4.5) contributes
$\sum_{j=1}^{\infty}A_{j,\pm}(s)$, having defined
\begin{equation}
A_{j,\pm}(s) \equiv 
\sum_{n=3}^{\infty} n^{-(2s+j-2)}
{(\sin \pi s)\over \pi}
\int_{0}^{\infty}dz \; z^{-2s}
{\partial \over \partial z} T_{j,\pm}(\tau(z)),
\label{(4.10)}
\end{equation}
where, from the formulae
\begin{equation}
T_{1,\pm}=p_{1,\pm},
\label{(4.11)}
\end{equation}
\begin{equation}
T_{2,\pm}=p_{2,\pm}-{1\over 2}p_{1,\pm}^{2},
\label{(4.12)}
\end{equation}
\begin{equation}
T_{3,\pm}=p_{3,\pm}-p_{1,\pm}p_{2,\pm}+{1\over 3}p_{1,\pm}^{3},
\label{(4.13)}
\end{equation}
we find
\begin{equation}
T_{1,\pm}=-{3\over 8}\tau \pm{1\over 2}\tau^{2}-{5\over 24}\tau^{3},
\label{(4.14)}
\end{equation}
\begin{equation}
T_{2,\pm}=-{3\over 16}\tau^{2} \pm{3\over 8}\tau^{3}
+{1\over 8}\tau^{4} \mp{5\over 8}\tau^{5}
+{5\over 16}\tau^{6}, 
\label{(4.15)}
\end{equation}
\begin{equation}
T_{3,\pm}=-{21\over 128}\tau^{3} \pm{3\over 8}\tau^{4}
+{509\over 640}\tau^{5} \mp{25\over 12}\tau^{6}
+{21\over 128}\tau^{7} \pm{15\over 8}\tau^{8}
-{1105 \over 1152}\tau^{9},
\label{(4.16)}
\end{equation}
and hence, in general,
\begin{equation}
T_{j,\pm}(\tau)=\sum_{a=j}^{3j}f_{a}^{(j,\pm)}\tau^{a}.
\label{(4.17)}
\end{equation}

We therefore find, from the first line of Eq. (3.14),
contributions to the generalized $\zeta$-function, from terms in
round brackets in Eq. (4.5), equal to
\begin{equation}
\delta \zeta_{A}^{\pm}(s)=\omega_{0}(s)F_{0}^{\pm}(s)+\sum_{j=1}^{\infty}
\omega_{j}(s)F_{j}^{\pm}(s),
\label{(4.18)}
\end{equation}
where, for all $\lambda=0,j$,
\begin{equation}
\omega_{\lambda}(s) \equiv \sum_{n=3}^{\infty}n^{-(2s+\lambda-2)}
= \zeta_{H}(2s+\lambda-2;3)
=\zeta_{R}(2s+\lambda-2)-1-2^{-(2s+\lambda-2)},
\label{(4.19)}
\end{equation}
while, from Eqs. (4.8)--(4.10),
\begin{equation}
F_{0}^{+}(s) \equiv {(\sin \pi s)\over \pi}\int_{0}^{\infty}
dz \; z^{-2s}{\partial \over \partial z}
\log \Bigr(1+(1+z^{2})^{-{1\over 2}}\Bigr),
\label{(4.20)}
\end{equation}
\begin{equation}
F_{0}^{-}(s) \equiv -2{(\sin \pi s)\over \pi}\int_{0}^{\infty}dz
\; {z^{-(2s-1)} \over (1+z^{2})}-F_{0}^{+}(s)
=-1-F_{0}^{+}(s),
\label{(4.21)}
\end{equation}
\begin{equation}
F_{j}^{\pm}(s) \equiv {(\sin \pi s)\over \pi}
\sum_{a=j}^{3j} L^{\pm}(s,a,0)f_{a}^{(j,\pm)},
\label{(4.22)}
\end{equation}
having set (this general definition will prove useful later, and arises
from a more general case, where $\tau^{a}$ is divided by the
$b$-th power of $(1 \pm \tau)$ in Eq. (4.17))
\begin{equation}
L^{\pm}(s,a,b) \equiv \int_{0}^{1}
\tau^{2s+a}(1-\tau)^{-s}(1+\tau)^{-s}
\Bigr(\pm b (1\pm \tau)^{-b-1}-a\tau^{-1}(1\pm \tau)^{-b}
\Bigr)d\tau.
\label{(4.23)}
\end{equation}

Moreover, on considering
\begin{equation}
L_{0}^{+}(s) \equiv {\pi \over \sin \pi s}F_{0}^{+}(s),
\label{(4.24)}
\end{equation}
and changing variable from $z$ to $\tau$ therein, all $L$-type
integrals above can be obtained from
\begin{equation}
Q(\alpha,\beta,\gamma) \equiv \int_{0}^{1}\tau^{\alpha}
(1-\tau)^{\beta}(1+\tau)^{\gamma}d\tau.
\label{(4.25)}
\end{equation}
In particular, we will need
\begin{equation}
L_{0}^{+}(s)=-Q(2s,-s,-s-1),
\label{(4.26)}
\end{equation}
\begin{equation}
L^{+}(s,a,b)=bQ(2s+a,-s,-s-b-1)-aQ(2s+a-1,-s,-s-b),
\label{(4.27)}
\end{equation}
where, from the integral representation of the
hypergeometric function, one has \cite{[17]}
\begin{equation}
Q(\alpha,\beta,\gamma)={\Gamma(\alpha+1)\Gamma(\beta+1)\over
\Gamma(\alpha+\beta+2)}F(-\gamma,\alpha+1;\alpha+\beta+2;-1).
\label{(4.28)}
\end{equation}
For example, explicitly, 
\begin{equation}
L_{0}^{+}(s)=-{\Gamma(2s+1)\Gamma(1-s)\over \Gamma(s+2)}
F(s+1,2s+1;s+2;-1).
\label{(4.29)}
\end{equation}

Now we exploit Eqs. (4.4), (4.5) and (4.18) to write
\begin{equation}
\zeta_{A}^{+}(s)=\delta \zeta_{A}^{+}(s)+{(\sin \pi s)\over \pi}
\sum_{n=3}^{\infty}n^{-(2s-2)}\int_{0}^{\infty}dz 
\left[{z^{-(2s-1)}\over 2(1+z^{2})}
+nz^{-(2s+1)} \left(\sqrt{1+z^{2}}-1 \right) \right].
\label{(4.30)}
\end{equation}
Hence we find
\begin{equation}
\zeta_{A}^{+}(0)=\lim_{s \to 0}\left[\omega_{0}(s)F_{0}^{+}(s)
+\sum_{j=1}^{\infty}\omega_{j}(s)F_{j}^{+}(s)
+\Bigr(\zeta_{A}^{+}(s)-\delta \zeta_{A}^{+}(s)\Bigr)\right].
\label{(4.31)}
\end{equation}
The first limit in Eq. (4.31) is immediately obtained 
by noting that
\begin{equation}
\lim_{s \to 0}
L_{0}^{+}(s)=-\log(2),
\label{(4.32)}
\end{equation}
and hence
\begin{equation}
\lim_{s \to 0}\omega_{0}(s)F_{0}^{+}(s)
=\lim_{s \to 0}\left[\zeta_{H}(2s-2;3){(\sin \pi s)\over \pi}
L_{0}^{+}(s)\right]=0.
\label{(4.33)}
\end{equation}
To evaluate the second limit in Eq. (4.31), we use
\begin{equation}
\lim_{s \to 0}L^{+}(s,a,0)=-1,
\label{(4.34)}
\end{equation}
and bear in mind  
that $\omega_{j}(s)$ is a meromorphic function with 
first-order pole, as $s \rightarrow 0$, 
only at $j=3$ by virtue of the limit
\begin{equation}
\lim_{y \to 1}\left[\zeta_{R}(y)-{1\over (y-1)}\right]=\gamma.
\label{(4.35)}
\end{equation}
Hence we find (see coefficients in Eq. (4.16)) 
\begin{eqnarray}
\lim_{s \to 0}\sum_{j=1}^{\infty}\omega_{j}(s)F_{j}^{+}(s)
&=& \lim_{s \to 0}{(\sin \pi s)\over \pi}\sum_{j=1}^{\infty}
\omega_{j}(s)\left[\sum_{a=j}^{3j}L^{+}(s,a,0)f_{a}^{(j,+)}\right]
\nonumber \\
&=& -{1\over 2}\sum_{a=3}^{9}f_{a}^{(3,+)}=-{1\over 720},
\label{(4.36)}
\end{eqnarray}
while, from Eqs. (4.30) and (4.28),
\begin{eqnarray}
\lim_{s \to 0}\Bigr(\zeta_{A}^{+}(s)-\delta \zeta_{A}^{+}(s)\Bigr)
&=& \lim_{s \to 0}\left({1\over 4}\zeta_{H}(2s-2;3)
+{1\over 4 \sqrt{\pi}}{\Gamma \left(s-{1\over 2}\right)
\over \Gamma(s+1)} \zeta_{H}(2s-3;3)\right) \nonumber \\
&=& -{5\over 4}+{1079\over 240} .
\label{(4.37)}
\end{eqnarray}
From Eqs. (4.31)--(4.37) we therefore obtain
\begin{equation}
\zeta_{A}^{+}(0)=-{5\over 4}+{1079\over 240}
-{1\over 2}\sum_{a=3}^{9}f_{a}^{(3,+)}
={146 \over 45} .
\label{(4.38)}
\end{equation}

An analogous procedure leads to
\begin{equation}
\zeta_{A}^{-}(0)=-{5\over 4}+{1079\over 240}+5
-{1\over 2}\sum_{a=3}^{9}f_{a}^{(3,-)}
={757\over 90},
\label{(4.39)}
\end{equation}
where the $+5$ contribution results from Eq. (4.21) when exploited
in Eq. (4.18). These results
have been double-checked by using also the powerful 
analytic technique in Ref. \cite{[14]}.

\section{Further spectral asymptotics: elliptic and non-elliptic parts}

As a next step, the second line of Eq. (3.14) 
suggests considering $\zeta$-functions
having the integral representation (cf. Eq. (4.4))
\begin{eqnarray}
\zeta_{B}^{\pm}(s)& \equiv &{(\sin \pi s)\over \pi}\sum_{n=3}^{\infty}
n^{-(2s-2)} \nonumber \\
& \; & \int_{0}^{\infty}
dz \; z^{-2s}{\partial \over \partial z}{\rm log}
\left[z^{-\beta_{\pm}(n)}\left(znI_{n}'(zn)+\left({z^{2}n^{2}\over 2}
\pm n \right)I_{n}(zn)\right)\right].
\label{(5.1)}
\end{eqnarray}
To begin, we exploit again the uniform asymptotic 
expansion of modified
Bessel functions and their first derivatives 
to find (cf. Eq. (4.5))
\begin{equation}
znI_{n}'(zn)+\left({z^{2}n^{2}\over 2} \pm n \right)I_{n}(zn)
\sim {n^{2}\over 2 \sqrt{2 \pi n}}{e^{n \eta} \over \sqrt{\tau}}
\left({1\over \tau}-\tau \right) 
\left (1+ \sum_{k=1}^{\infty}{r_{k,\pm}(\tau)\over n^{k}}
\right),
\label{(5.2)}
\end{equation}
where we have (bearing in mind that $u_{0}=v_{0}=1$)
\begin{equation}
r_{k,\pm}(\tau) \equiv u_{k}(\tau)
+{2\tau \over (1-\tau^{2})}
\Bigr((v_{k-1}(\tau) \pm \tau 
u_{k-1}(\tau) \Bigr),
\label{(5.3)}
\end{equation}
for all $k \geq 1$. Hereafter we set
\begin{equation}
\Omega \equiv \sum_{k=1}^{\infty}{r_{k,\pm}(\tau(z))\over n^{k}},
\label{(5.4)}
\end{equation}
and rely upon the formula
\begin{equation}
\log(1+\Omega) \sim \sum_{k=1}^{\infty}(-1)^{k+1}
{\Omega^{k}\over k}
\label{(5.5)}
\end{equation}
to evaluate the uniform asymptotic expansion (cf. Eq. (4.7))
\begin{equation}
{\rm log}\left(1+\sum_{k=1}^{\infty}{r_{k,\pm}(\tau(z))\over n^{k}}
\right) \sim \sum_{k=1}^{\infty}{R_{k,\pm}(\tau(z))\over n^{k}}.
\label{(5.6)}
\end{equation}
The formulae yielding $R_{k,\pm}$ from $r_{k,\pm}$ are exactly as in
Eqs. (4.11)--(4.13), with $T$ replaced by $R$ and $p$ replaced by $r$
(see, however, comments below Eq. (5.10)).
Hence we find, bearing in mind Eq. (5.3),
\begin{equation}
R_{1,\pm}=(1 \mp \tau)^{-1}
\left({17\over 8}\tau \mp{1\over 8}\tau^{2}-{5\over 24}\tau^{3}
\pm{5\over 24}\tau^{4}\right),
\label{(5.7)}
\end{equation}
\begin{equation}
R_{2,\pm}=(1 \mp \tau)^{-2}
\left(-{47\over 16}\tau^{2} \pm{15\over 8}\tau^{3}-{21\over 16}\tau^{4}
\pm{3\over 4}\tau^{5}-{1\over 16}\tau^{6} \mp{5\over 8}\tau^{7}
+{5\over 16}\tau^{8}\right),
\label{(5.8)}
\end{equation}
\begin{eqnarray}
R_{3,\pm}&=& (1 \mp \tau)^{-3}\biggr(
{1721\over 384}\tau^{3} \mp{441\over 128}\tau^{4}+{597\over 320}\tau^{5}
\mp{1033\over 960}\tau^{6} 
+{239\over 80}\tau^{7} \nonumber \\
&\mp & {28\over 5}\tau^{8}
+{2431\over 576}\tau^{9} \pm{221\over 192}\tau^{10} 
- {1105\over 384}\tau^{11} \pm{1105\over 1152}\tau^{12}\biggr),
\label{(5.9)}
\end{eqnarray}
and therefore
\begin{equation}
R_{j,\pm}(\tau(z))=(1 \mp \tau)^{-j}\sum_{a=j}^{4j}C_{a}^{(j,\pm)}
\tau^{a},
\label{(5.10)}
\end{equation}
where, unlike what happens for the $T_{j,\pm}$ polynomials, 
the exponent of $(1 \mp \tau)$ never vanishes. Note that, at
$\tau=1$ (i.e. $z=0$), our $r_{k,+}(\tau)$ and $R_{k,+}(\tau)$
are singular. Such a behaviour
is not seen for any of the strongly elliptic boundary-value
problems (see third item in Ref. \cite{[11]}). This technical 
difficulty motivates our efforts below and is interpreted by us as 
a clear indication of the lack of strong ellipticity proved,
on general ground, in Ref. \cite{[6]}.

The $\zeta_{B}^{-}(s)$ function is more easily dealt with.
It indeed receives contributions from terms in 
round brackets in Eq. (5.2) equal to (cf. Eq. (4.9) and
bear in mind that $\beta_{-}-\beta_{+}=2$ in Eq. (5.1))
\begin{eqnarray}
B_{-}(s)& \equiv & \sum_{n=3}^{\infty}n^{-(2s-2)}{(\sin \pi s)\over \pi}
\int_{0}^{\infty}dz \; z^{-2s}{\partial \over \partial z}
\log \left({{1\over \tau(z)}-\tau(z) \over z^{2}}\right)
\nonumber \\
&=& \omega_{0}(s){(\sin \pi s)\over \pi}
\int_{0}^{\infty}dz \; z^{-2s}{\partial \over \partial z}
\log {1\over \sqrt{1+z^{2}}}
=-{1\over 2}\omega_{0}(s),
\label{(5.11)}
\end{eqnarray}
and $\sum_{j=1}^{\infty}B_{j,-}(s)$, having defined 
(cf. Eq. (4.10))
\begin{equation}
B_{j,-}(s) \equiv \omega_{j}(s){(\sin \pi s)\over \pi}
\int_{0}^{\infty}dz \; z^{-2s}{\partial \over \partial z}
R_{j,-}(\tau(z))
=\omega_{j}(s){(\sin \pi s)\over \pi}\sum_{a=j}^{4j} 
L^{+}(s,a,j)C_{a}^{(j,-)}.
\label{(5.12)}
\end{equation}
On using the same method as in Sec. IV, 
the formulae (5.1)--(5.12) lead to (we find 
$L^{+}(0,a,3)=-{1\over 8}$, independent of $a$, below)
\begin{equation}
\zeta_{B}^{-}(0)=-{5\over 4}+{1079\over 240}+{5\over 2}
+{1\over 2}\sum_{a=3}^{12}L^{+}(0,a,3)C_{a}^{(3,-)} 
={206\over 45},
\label{(5.13)}
\end{equation}
a result which agrees with a derivation of $\zeta_{B}^{-}(0)$ 
relying upon the method of Ref. \cite{[14]}.

Although we have stressed after Eq. (5.10) 
the problems with the $\zeta_{B}^{+}(s)$
part, for the moment let us proceed formally in the same way
as above. Thus we define, in analogy to Eq. (5.11),
\begin{equation}
B_{+}(s)\equiv \omega_{0}(s){(\sin \pi s)\over \pi}
\int_{0}^{\infty}dz \; z^{-2s}{\partial \over \partial z}
\log \left({1\over \tau(z)}
-\tau(z)\right),
\label{(5.14)}
\end{equation}
and, in analogy to Eq. (5.12),
\begin{equation}
B_{j,+}(s) \equiv \omega_{j}(s){(\sin \pi s)\over \pi}
\int_{0}^{\infty}dz \; z^{-2s}{\partial \over \partial z}
R_{j,+}(\tau(z)).
\label{(5.15)}
\end{equation}
In order to make the presentation as transparent as possible, we
write out the derivatives of $R_{j,+}$. On changing integration
variable from $z$ to $\tau$ we define
\begin{equation}
C_{j}(\tau) \equiv {\partial \over \partial \tau}R_{j,+}(\tau),
\label{(5.16)}
\end{equation}
and we find the following results:
\begin{equation}
C_{1}(\tau)=
(1-\tau)^{-2}\left({17\over 8}-{1\over 4}\tau-{1\over 2}\tau^{2}
+{5\over 4}\tau^{3}-{5\over 8}\tau^{4}\right),
\label{(5.17)}
\end{equation}
\begin{equation}
C_{2}(\tau)=
(1-\tau)^{-3}\left(-{47\over 8}\tau+{45\over 8}\tau^{2}
-{57\over 8}\tau^{3}+{51\over 8}\tau^{4}-{21\over 8}\tau^{5}
-{33\over 8}\tau^{6}+{45\over 8}\tau^{7}-{15\over 8}\tau^{8}
\right),
\label{(5.18)}
\end{equation}
\begin{eqnarray}
C_{3}(\tau)&=&
(1-\tau)^{-4}\biggr({1721\over 128}\tau^{2}-{441\over 32}\tau^{3}
+{1635\over 128}\tau^{4}-{163\over 16}\tau^{5}
+{1545\over 64}\tau^{6}-{227\over 4}\tau^{7} \nonumber \\
&+& {4223\over 64}\tau^{8}-{221\over 16}\tau^{9}
-{5083\over 128}\tau^{10}+{1105\over 32}\tau^{11}
-{1105\over 128}\tau^{12}\biggr),
\label{(5.19)}
\end{eqnarray}
so that the general expression of $C_{j}(\tau)$ reads as
\begin{equation}
C_{j}(\tau)=
\sum_{a=j-1}^{4j}
{K_{a}^{(j)}\tau^{a}\over (1-\tau)^{j+1}}, \;
\forall j =1,{\ldots} , \infty \; .
\label{(5.20)}
\end{equation}
These formulae engender a $\zeta_{B}^{+}(0)$ which can be defined,
after change of variable from $z$ to $\tau$,
by splitting the integral with respect to $\tau$, in the integral
representation of $\zeta_{B}^{+}(s)$, according to the identity
$$
\int_{0}^{1}d\tau=\int_{0}^{\mu}d\tau+\int_{\mu}^{1}d\tau,
$$
and taking the limit as $\mu \rightarrow 1$ {\it after having evaluated 
the integral}. More precisely, since the integral on the left-hand side
is independent of $\mu$, we can choose $\mu$ small on the right-hand
side so that, in the interval $[0,\mu]$
(and only there!), we can use the uniform 
asymptotic expansion of the integrand where the negative powers of
$(1-\tau)$ are harmless. Moreover, independence of $\mu$ also
implies that, after having evaluated the integrals on the right-hand
side, we can take the $\mu \rightarrow 1$ limit. Within this framework,
the limit as $\mu \rightarrow 1$ of the second integral on the
right-hand side yields vanishing contribution to the asymptotic
expansion of $\zeta_{B}^{+}(s)$.

With this {\it caveat}, on defining (cf. Eq. (4.25))
\begin{equation}
Q_{\mu}(\alpha,\beta,\gamma) \equiv \int_{0}^{\mu}
\tau^{\alpha}(1-\tau)^{\beta}(1+\tau)^{\gamma} d\tau,
\label{(5.21)}
\end{equation}
we obtain the representations
\begin{equation}
B_{+}(s)=-\omega_{0}(s){(\sin \pi s)\over \pi}\Bigr[
-Q_{\mu}(2s,-s-1,-s)+Q_{\mu}(2s,-s,-s-1)
-Q_{\mu}(2s-1,-s,-s)\Bigr],
\label{(5.22)}
\end{equation}
\begin{equation}
B_{j,+}(s)=-\omega_{j}(s){(\sin \pi s)\over \pi}
\sum_{a=j-1}^{4j}K_{a}^{(j)}Q_{\mu}(2s+a,-s-j-1,-s).
\label{(5.23)}
\end{equation}
The relevant properties of $Q_{\mu}(\alpha,\beta,\gamma)$ can be
obtained by observing that this function is nothing but a
hypergeometric function of two variables \cite{[17]}, i.e.
\begin{equation}
Q_{\mu}(\alpha,\beta,\gamma)={\mu^{\alpha+1}\over \alpha+1}
F_{1}(\alpha+1,-\beta,-\gamma,\alpha+2;\mu,-\mu).
\label{(5.24)}
\end{equation}
In detail, a summary of results needed to consider the limiting 
behaviour of $\zeta_{B}^{+}(s)$ as $s \rightarrow 0$ is
\begin{equation}
\omega_{0}(s){(\sin \pi s)\over \pi} \sim -5s+
{\rm O}(s^{2}),
\label{(5.25)}
\end{equation}
\begin{equation}
\omega_{j}(s){(\sin \pi s)\over \pi} \sim
{1\over 2}\delta_{j,3}+{\tilde b}_{j,1}s+{\rm O}(s^{2}),
\label{(5.26)}
\end{equation}
\begin{equation}
\lim_{\mu \to 1}Q_{\mu}(2s,-s-1,-s) \sim -{1\over s}+{\rm O}(s^{0}),
\label{(5.27)}
\end{equation}
\begin{equation}
\lim_{\mu \to 1}Q_{\mu}(2s,-s,-s-1) \sim \log(2)+{\rm O}(s),
\label{(5.28)}
\end{equation}
\begin{equation}
\lim_{\mu \to 1}Q_{\mu}(2s-1,-s,-s) \sim {1\over 2s}
+{\rm O}(s),
\label{(5.29)}
\end{equation}
\begin{eqnarray}
\lim_{\mu \to 1}Q_{\mu}(2s+a,-s-j-1,-s) 
&=&{\Gamma(-j-s)\Gamma(a+2s+1)\over \Gamma(a-j+s+1)}
F(a+2s+1,s,a-j+s+1;-1) \nonumber \\
&\sim &  {b_{j,-1}(a)\over s}+b_{j,0}(a)+{\rm O}(s),
\label{(5.30)}
\end{eqnarray}
where 
\begin{equation}
{\tilde b}_{j,1}=-1-2^{2-j}+\zeta_{R}(j-2)(1-\delta_{j,3})
+\gamma \delta_{j,3},
\label{(5.31)}
\end{equation}
\begin{equation}
b_{j,-1}(a)= {(-1)^{j+1}\over j!}
{\Gamma(a+1)\over \Gamma(a-j+1)}(1-\delta_{a,j-1}),
\label{(5.32)}
\end{equation}
and we only strictly need $b_{3,0}(a)$ which,
unlike the elliptic cases studied earlier,
now depends explicitly on $a$. In our case, we find
$$
b_{3,0}(2)={1\over 24}, \; 
b_{3,0}(3)={2\over 3}-\log(2), \;
b_{3,0}(4)={67\over 24}-4 \log(2), \;
b_{3,0}(5)={95\over 12}-10 \log(2), 
$$
$$
b_{3,0}(6)={143\over 8}-20 \log(2), \;
b_{3,0}(7)={139\over 4}-35 \log(2), \;
b_{3,0}(8)={2433\over 40}-56 \log(2), 
$$
$$
b_{3,0}(9)={1971\over 20}-84 \log(2), \;
b_{3,0}(10)={8429\over 56}-120 \log(2), 
$$
$$
b_{3,0}(11)={12289 \over 56}-165 \log(2), \;
b_{3,0}(12)={155315\over 504}-220 \log(2).
$$

Remarkably, the coefficient of ${1\over s}$ in the small-$s$
behaviour of the generalized 
$\zeta$-function $\zeta_{B}^{+}(s)$ is zero because it is equal to
$$
\sum_{a=2}^{12}b_{3,-1}(a)K_{a}^{(3)}= {1\over 6}
\sum_{a=3}^{12}a(a-1)(a-2)K_{a}^{(3)},
$$
which vanishes by virtue of the rather peculiar general property
\begin{equation}
\sum_{a=j}^{4j}{\Gamma(a+1)\over \Gamma(a-j+1)}K_{a}^{(j)}
=\sum_{a=j}^{4j}\prod_{l=0}^{j-1}(a-l)K_{a}^{(j)}=0, \;
\forall j=1,{\ldots} ,\infty,
\label{(5.33)}
\end{equation}
and hence we find eventually
\begin{eqnarray}
\zeta_{B}^{+}(0)&=& -{5\over 4}+{1079\over 240}
+{5\over 2}-{1\over 2}\sum_{a=2}^{12}b_{3,0}(a)K_{a}^{(3)}
-\sum_{j=1}^{\infty}{\tilde b}_{j,1}
\sum_{a=j-1}^{4j}b_{j,-1}(a)K_{a}^{(j)} \nonumber \\
&=& {5\over 4}+{1079\over 240}+{599\over 720}
={296\over 45},
\label{(5.34)}
\end{eqnarray}
because the infinite sum on the first line of Eq. (5.34) vanishes 
by virtue of Eqs. (5.32) and (5.33), and exact cancellation of
$\log(2)$ terms is found to occur.

To cross-check our analysis we remark that,
on applying the technique of Ref. \cite{[14]}, one finds
\begin{equation}
\zeta_{B}^{+}(0)=-{15\over 4}+{1079\over 240}-{1\over 720}
={67\over 90},
\label{(5.35)}
\end{equation}
where $-{1\over 720}$ results from working in the 
$n \rightarrow \infty$ and $w \rightarrow 0$ limit in 
$$
\left(I_{n}'(w)+\left({w\over 2}+{n\over w}\right)I_{n}(w)\right)
$$
on the second line of Eq. (3.14); such a term then reduces to
$\left(I_{n}'(w)+{n\over w}I_{n}(w)\right)$.
A possible interpretation of the discrepancy between (5.34) and
(5.35) is that, when strong ellipticity is violated, prescriptions
for defining a $\zeta(0)$ value exist but are inequivalent.

Remaining contributions to $\zeta(0)$,
being obtained from strongly elliptic sectors of the
boundary-value problem, are instead found to agree with the
results in Ref. \cite{[7]}, i.e.
\begin{equation}
\zeta(0)[{\rm transverse} \; 
{\rm traceless} \; {\rm modes}]=-{278\over 45},
\label{(5.36)}
\end{equation}
\begin{equation}
\zeta(0)[{\rm coupled} \; {\rm vector} 
\; {\rm modes}]={494\over 45},
\label{(5.37)}
\end{equation}
\begin{equation}
\zeta(0)[{\rm decoupled} \; {\rm vector} 
\; {\rm mode}]=-{15\over 2},
\label{(5.38)}
\end{equation}
\begin{equation}
\zeta(0)[{\rm scalar} \; {\rm modes}(a_{1},e_{1};a_{2},b_{2},e_{2})]
=-17,
\label{(5.39)}
\end{equation}
\begin{equation}
\zeta(0)[{\rm scalar} \; {\rm ghost} 
\; {\rm modes}]=-{149\over 45},
\label{(5.40)}
\end{equation}
\begin{equation}
\zeta(0)[{\rm vector} \; {\rm ghost} 
\; {\rm modes}]={77\over 90},
\label{(5.41)}
\end{equation}
\begin{equation}
\zeta(0)[{\rm decoupled} \; {\rm ghost} \; {\rm mode}]={5\over 2}.
\label{(5.42)}
\end{equation}

\section{Concluding remarks}

We have obtained the
analytically continued eigenvalue conditions for metric perturbations
on the Euclidean four-ball, in the presence of boundary conditions 
completely invariant under infinitesimal diffeomorphisms in the de
Donder gauge and with $\alpha$ parameter set to $1$ in Eq. (1.10). 
Second, this has made it possible to prove, for the first time
in the literature, that only one sector of the scalar-mode
determinant is responsible for lack of strong ellipticity of the
boundary-value problem (see second line of Eq. (3.14) and the analysis
in Secs. IV and V). The first novelty with respect to the work in
Ref. \cite{[6]} is a better understanding of the elliptic and
non-elliptic sectors of spectral asymptotics for Euclidean 
quantum gravity. Moreover, as far as we know, the detailed spectral
asymptotics for $\zeta$-functions of Secs. IV and V was missing
in the literature. We have also shown that one can indeed obtain a
regular $\zeta$-function asymptotics at small $s$ in
the non-elliptic case by virtue of the
remarkable identity (5.33), here obtained for the first time.
Our prescription for the
$\zeta(0)$ value differs from  
the result first obtained in
Ref. \cite{[7]}, where, however, neither the strong ellipticity
issue \cite{[6]} nor the non-standard spectral asymptotics of
our Sec. V had been considered.

From the point of view of general formalism of Euclidean quantum
gravity, three alternative pictures seem therefore to emerge:
\vskip 0.3cm
\noindent
(i) The remarkable factorization of eigenvalue conditions, with
resulting isolation of elliptic part of spectral asymptotics
(transverse-traceless, vector and ghost modes, all modes in
finite-dimensional sub-spaces and three of the four equations for
scalar modes), suggests trying to re-assess functional integrals on
manifolds with boundary, with the hope of being able to obtain
unique results from the
non-elliptic contribution. If this cannot be achieved, 
the two alternatives below 
should be considered again.
\vskip 0.3cm
\noindent
(ii) Luckock boundary conditions \cite{[18]},
which engender BRST-invariant amplitudes but are not 
diffeomorphism invariant \cite{[3]}. They have already been applied
by Moss and Poletti \cite{[19]}, \cite{[20]}.
\vskip 0.3cm
\noindent
(iii) Non-local boundary conditions that lead to
surface states in quantum cosmology and pseudo-differential 
operators on metric and ghost modes \cite{[21]}. Surface states are
particularly interesting since they describe a transition from quantum
to classical regime in cosmology entirely ruled by the strong ellipticity
requirement, while pseudo-differential operators 
are a source of technical complications.

There is therefore encouraging evidence in favour of Euclidean 
quantum gravity being able to drive further developments in
quantum field theory, quantum cosmology and spectral asymptotics
(see early mathematical papers in Refs. \cite{[22]}, \cite{[23]})
in the years to come.

\appendix
\section{Strong ellipticity}
For an operator of Laplace type, the boundary-value problem is
strongly elliptic with respect to the cone ${\bf C}-{\bf R}_{+}$
if, for any cotangent vector $u$ on the boundary $\cal B$,
for any $\lambda \in {\bf C}-{\bf R}_{+}$, for any pair
$(u,\lambda) \not = (0,0)$, there exists a unique solution $\varphi$
of the differential equation ($r$ being the geodesic distance to
the boundary $\cal B$)
\begin{equation}
\left[-{\partial^{2}\over \partial r^{2}}+u_{k}u^{k}
-\lambda \right]\varphi(r)=0,
\label{(A1)}
\end{equation}
subject to the asymptotic condition
\begin{equation}
\lim_{r \to \infty}\varphi(r)=0,
\label{(A2)}
\end{equation}
and to the boundary conditions (here $\varphi(r)=\chi e^{-\sigma r}$
with $\sigma \equiv \sqrt{u_{k}u^{k}-\lambda}$)
\begin{equation}
\pi \varphi(r=0)=\psi_{0}, \; 
iT \varphi(r=0)+(I-\pi)\varphi'(r=0)=\psi_{1},
\label{(A3)}
\end{equation}
where $\pi$ is the same projector as in Eq. (1.1),
$iT$ is the leading symbol of that part of the boundary 
operator which involves tangential derivatives,
while $\psi_{0}$ and $\psi_{1}$ are arbitrary boundary data. 
Eventually, all this is equivalent to proving positivity of the
matrix $I \sqrt{u_{k}u^{k}}-iT$ \cite{[6]}.

\section{Bessel functions}
In Sec. II we exploit the following identities obeyed by modified
Bessel functions of first kind:
\begin{equation}
I_{n+1}(w)=I_{n}'(w)-{n \over w}I_{n}(w),
\label{(B1)}
\end{equation}
\begin{equation}
I_{n-1}(w)=I_{n}'(w)+{n\over w}I_{n}(w),
\label{(B2)}
\end{equation}
\begin{equation}
I_{n+2}(w)=\left(1+{2n(n+1)\over w^{2}}\right)I_{n}(w)
-{2(n+1)\over w}I_{n}'(w),
\label{(B3)}
\end{equation}
\begin{equation}
I_{n-2}(w)=\left(1+{2n(n-1)\over w^{2}}\right)I_{n}(w)
+{2(n-1)\over w}I_{n}'(w),
\label{(B4)}
\end{equation}
\begin{equation}
I_{n+2}'(w)=-{2(n+1)\over w}\left(1+{n(n+2)\over w^{2}}\right)I_{n}(w)
+\left(1+{2(n+1)(n+2)\over w^{2}}\right)I_{n}'(w),
\label{(B5)}
\end{equation}
\begin{equation}
I_{n-2}'(w)={2(n-1)\over w}\left(1+{n(n-2)\over w^{2}}\right)I_{n}(w)
+\left(1+{2(n-1)(n-2)\over w^{2}}\right)I_{n}'(w).
\label{(B6)}
\end{equation}

In Secs. IV and V we use the uniform asymptotic expansion of modified
Bessel functions $I_{\nu}$ first found by Olver \cite{[24]}:
\begin{equation}
I_{\nu}(z\nu) \sim {e^{\nu \eta}\over \sqrt{2\pi \nu}
(1+z^{2})^{1\over 4}}\left(1+\sum_{k=1}^{\infty}
{u_{k}(\tau)\over \nu^{k}}\right),
\label{(B7)}
\end{equation}
where
\begin{equation}
\tau \equiv (1+z^{2})^{-{1\over 2}}, \;
\eta \equiv (1+z^{2})^{1\over 2}
+{\rm log}\left({z \over 1+\sqrt{1+z^{2}}}\right).
\label{(B8)}
\end{equation}
This holds for $\nu \rightarrow \infty$ at fixed $z$. The polynomials
$u_{k}(\tau)$ can be found from the recurrence relation \cite{[15]}
\begin{equation}
u_{k+1}(\tau)={1\over 2}\tau^{2}(1-\tau^{2})u_{k}'(\tau)
+{1\over 8}\int_{0}^{\tau}d\rho \; (1-5\rho^{2})u_{k}(\rho),
\label{(B9)}
\end{equation}
starting with $u_{0}(\tau)=1$. Moreover, the first derivative of
$I_{\nu}$ has the following uniform asymptotic expansion at large $\nu$ 
and fixed $z$:
\begin{equation}
I_{\nu}'(z \nu) \sim {e^{\nu \eta}\over \sqrt{2 \pi \nu}}
{(1+z^{2})^{1\over 4}\over z}\left(1+\sum_{k=1}^{\infty}
{v_{k}(\tau)\over \nu^{k}}\right),
\label{(B10)}
\end{equation}
with the $v_{k}$ polynomials determined from the $u_{k}$ according to
\cite{[15]}
\begin{equation}
v_{k}(\tau)=u_{k}(\tau)+\tau(\tau^{2}-1)\left[
{1\over 2}u_{k-1}(\tau)+\tau u_{k-1}'(\tau)\right],
\label{(B11)}
\end{equation}
starting with $v_{0}(\tau)=u_{0}(\tau)=1$.

\acknowledgments
The present paper is dedicated to the loving memory 
of Mirella Russo and Bryce DeWitt.
We are indebted to Ivan Avramidi for inspiration provided by
previous collaboration with some of us and by continuous
correspondence, and to Gerd Grubb for enlightening correspondence.
K. Kirsten is grateful to the Baylor University Summer Sabbatical
Program and to the INFN for financial support. The 
work of G. Esposito and K. Kirsten has been partially supported
also by PRIN 2002 {\it SINTESI}.

\end{document}